\RequirePackage{docswitch}

\documentclass[twocolumn]{aastex61}

\usepackage{lsstdesc_macros}
\usepackage{graphicx}
\graphicspath{{./}{./figures/}}

\usepackage{listings}
\usepackage{amsmath}
\definecolor{codegreen}{rgb}{0,0.6,0}

\newcommand{\nsize}{20000}

\newcommand{\fillfactor}{$90 \%$~}
\newcommand{\healpix}{\texttt{HEALPix}}
\newcommand{\transients}{transients }
\newcommand{\snana}{{\texttt{SNANA}}}

\newcommand{\opsim}{\texttt{OpSim}}
\newcommand{\oss}{\texttt{OpSimSummary}}
\newcommand{\lsst}{\texttt{LSST}}
\newcommand{\plasticc}{PLAsTiCC}
\newcommand{\lsststart}{2022}
\newcommand{\tdas}{TDAS}
\usepackage[normalem]{ulem}

\usepackage{lineno}
\usepackage{url}

\begin{document}

\title{Enabling Catalog Simulations of Transient and Variable Sources based on LSST Cadence Strategies}

\author{Rahul~Biswas}
\affiliation{The Oskar Klein Centre for CosmoParticle Physics, Department of Physics, Stockholm University, AlbaNova, Stockholm SE-10691}
\affiliation{The eScience Institute, University of Washington, Seattle, WA 98195, USA}
\affiliation{Department of Astronomy, University of Washington, Seattle, WA 98195, USA}
\author{Scott~F.~Daniel}
\affiliation{Department of Astronomy, University of Washington, Seattle, WA 98195, USA}
\author{R~Hlo\v{z}ek}
\affiliation{Department of Astronomy and Astrophysics, University of Toronto, ON M5S 3H4, Canada}
\affiliation{Dunlap Institute of Astronomy and Astrophysics, University of Toronto, ON M5S 3H4, Canada}
\author{A.~G.~Kim}
\affiliation{Physics Division, Lawrence Berkeley National Laboratory, 1 Cyclotron Road, Berkeley, CA 94720 USA }
\author{Peter~Yoachim}
\affiliation{Department of Astronomy, University of Washington, Seattle, WA 98195, USA}

\collaboration{(LSST Dark Energy Science Collaboration)}

\begin{abstract}
The Large Synoptic Survey Telescope ({\lsst}) project will conduct a ten year multi-band survey starting in {\lsststart}. Observing strategies for this survey are being actively investigated, and the science capabilities can be best forecasted on the basis of simulated strategies from the {\lsst} Operations Simulator ({\opsim}). {\opsim} simulates a stochastic realization of the sequence of {\lsst} pointings over the survey duration, and is based on a model of the observatory (including telescope) and historical data of observational conditions. {\opsim} outputs contain a record of each simulated pointing of the survey along with a complete characterization of the pointing in terms of observing conditions, and some useful quantities derived from the characteristics of the pointing. Thus, each record can be efficiently used to derive the properties of observations of all astrophysical sources found in that pointing. However, in order to obtain the time series of observations (light curves) of a set of sources, it is often more convenient to compute all observations of an astrophysical source, and iterate over sources. In this document, we describe the open source python package {\oss} which allows for a convenient reordering. The objectives of this package are to provide users with an Application Programming Interface (API) for accessing all such observations and summarizing this information in the form intermediate data products usable by third party software such as {\snana}, thereby also bridging the gap between official {\lsst} products and pre-existing simulation codes. 
\end{abstract}



\section{Introduction}
\label{sec:intro}
The Large Synoptic Survey Telescope ({\lsst}) project will conduct a multi-band imaging survey~\citep{ScienceBook}~ of the sky with a 3.2 gigapixel camera on a 8 m class ground based telescope at Cerro Pachon, Chile with a field of view of about 10 square degrees. The survey is scheduled to start taking data for science operations in {\lsststart}, and cover most of the Southern sky to median single visit depths of $ \rm{r} \sim 24.3,$ revisiting each location frequently. The combination of large sky coverage, high depth and repeated visits enables several major scientific goals such as studying the Solar System, astrophysical transients and variables, the Milky Way, and the physics of dark matter and dark energy~\citep{overview}. The efficacy of such investigations, particularly the Time Domain Astronomy programs involving observations of Time Dependent Astronomical Sources ({\tdas}) such as transients, variable stars, AGN, as well as solar system objects depends critically on the observing strategy used to determine the sequence of pointings of the telescope.

Forecasting the performance of a science program with {\lsst} survey strategies through the analysis of mock catalogs of observations of sources relevant to the science program is important and timely.
Such forecasts are essential for the study of the impact of survey design and strategy. They are also instrumental in developing and testing appropriate analysis methods. Simulation of such mock catalog requires models of the astrophysical sources, models of the observing instrument and analysis methods used to reduce the real data to such catalogs, and a model of the survey strategy along with a model of the observing conditions.  

During the survey, the {\lsst} project will make observations of the sky, by pointing in different directions, recording the image for a certain amount of time and then processing the image. This procedure of procuring an image of a sky location for processing is referred to as a `visit' in the {\lsst} literature, and the visit itself may involve two `snaps' separated by the shutter closing (current baseline strategies have two snaps of 15 seconds each resulting in a visit of exposure of 30 seconds). A visit will be followed by a possible slew of the telescope to a different location, after which a new visit starts again to repeat the cycle. As each visit is short,
the observing conditions determined by the atmospheric and sky conditions can be approximated as constant during a visit. Currently, the {\lsst} project simulates observations during its survey period using the Operations Simulator ({\opsim}) \citep{2014SPIE.9150E..15D,2016SPIE.9910E..13D,2016SPIE.9911E..25R}. This is done with a prototype scheduler queuing visits according to a strategy designed to optimize science using a high fidelity model of the telescope to calculate times required for telescope slews, and real time observing conditions simulated using an empirical model of the sky and atmosphere. 
The output of such an {\opsim} simulation is a sequence of all the visits during the survey, and includes quantities required to describe the state of the telescope after each visit, and the observing conditions during the pointing. Such {\opsim} outputs may be considered realized forecasts of {\lsst}.

Such forecasts of science performance can be done in several ways representing different trade-offs between computational/storage costs and the level of accuracy. On the low resource end, the Metric Analysis Framework~\citep[MAF]{2014SPIE.9149E..0BJ} uses `metrics' which are proxies of the scientific performance of the survey. Such proxies are built as functions of quantities related to observational conditions, and are usually designed by scientists on the basis of past experiences and intuition. Such metrics are extremely useful for studying the impact of survey strategy.
 On the resource intensive end, there are image simulation codes (PhoSim~\citep{2015ApJS..218...14P} and ImSim~{\footnote{\url{https://github.com/LSSTDESC/imSim}}}) capable of using the {\opsim} outputs and producing detailed realistic simulations of {\lsst} images, but are computationally expensive in terms of generation and storage. Further, analysis of these images follows the expected {\lsst} image processing using the {\lsst} software stack~\citep{2015arXiv151207914J} and therefore best represents the scientific performance of {\lsst}. However, this analysis is also resource intensive, leading to the conclusion that such end-to-end explorations are hard, and therefore can only be used in a limited number of cases. An interesting middle ground is provided by catalog simulations which utilize the {\opsim} outputs to obtain the properties of visits, models of the astrophysical sources obtained from previous data or theoretical calculations, and models of aspects of the image processing procedure in the {\lsst} analyses. These simulated catalogs are mock realizations of the information contained in {\lsst} data releases (DRP) containing forced photometry of all time dependent objects detected by the {\lsst}, expected to be released through a (nearly) annual frequency~\citep{DPDD}, replacing the step of image analysis and reduction to catalog by an assumed model (which can in turn be improved through studies involving reprocessing older data and image simulations). 

For the more abundant categories of time dependent sources such as Type Ia Supernovae (SNIa), it is critical for catalog simulations to use distributed computing to speed up the simulations. There are at least two natural paradigms of organizing the distribution of compute resources. The first alternative (a) is to calculate the observed quantities corresponding to each telescope visit at a particular instance of time, which may be further split into smaller spatial regions. Indeed, this is almost essential for any image simulations, and is an approach utilized in generating `Instance Catalogs' by the {\lsst} Catalog Simulations (CatSim)~\citep{2010SPIE.7738E..1OC,2014SPIE.9150E..14C} that are used as intermediate data products by Image Simulation software like PhoSim and ImSim. These Instance Catalogs are catalogs of astrophysical objects in the simulated universe whose light is expected to impinge on the {\lsst} CCDs on that particular visit, along with a complete description of their astrophysical properties at that instance of time. In this method, obtaining the visit information is simple, however the state of the transient objects needs to be persisted from one visit to another, and the output of several visits have to be serialized before the light curves of the transients can be built. In the second approach (b) popular in the transient world, the paradigm involves distributing each astrophysical source (or groups thereof) to different resources, and simulating all of the observations of the source over a sequence of times. While this automatically leads to outputs with light curves for different objects in exactly the format useful for analysis, this calls for collecting the correct sequence of visits at a particular location, which is the only non-trivial step remaining.

Our objective in this work is to provide a solution to the collection of the correct sequence of visits for a transient or variable source to make alternative (b) simple. As described in the rest of the document, we do this by providing an open source package with a simple public Application Programming Interface (API) that users can use to obtain such sequences of visits. We also recognize that there are useful and often used codes like {\snana} ~\citep{2009PASP..121.1028K,2018arXiv181102379K} which are used to produce catalog simulations of time dependent sources, that demand specific forms of inputs aggregating this information. To enable the use of this code, we also provide a script which produces an intermediate data product (an observation library file in the {\snana}  terminology) in exactly the input form desired, so that this can work out of the box with {\snana}  simulations.

\section{Methods}
While we will not discuss the simulations of time dependent astronomical sources here, we start this section by noting the information about observations necessary for such simulations that are available from {\opsim} outputs, while a separate code (not provided in this work) is necessary to model the population of astrophysical objects themselves to get simulated observations.
In order to simulate catalogs of {\tdas}, one needs to simulate the observed 'flux' or photon counts of a source of known apparent brightness,
as parameterized by the specific flux $F_\nu(\lambda)$ at the top of the earth's atmosphere, and the uncertainty in the measured flux.  
The measured flux, or rather the counts of photons received from an astrophysical point source, or the sky are modelled as random variables that follow a Poisson distribution, where the expected counts from the source and the sky can be calculated from the physical parameters of the telescope and instruments, a knowledge of the effective point spread function (PSF), and the specific flux per unit area of the sky.
(see Appendix~\ref{appendix:snr} or ~\cite{LSE-40} for a more comprehensive discussion). 
 The expected counts of photons from astrophysical sources and the sky may be written (please see appendix.~\ref{appendix:snr} for a derivation, here we only use a summary of the results) in terms of the source magnitude and the sky brightness $m_{sky}$
\begin{equation}
     c_{source} = \kappa 10^{-0.4 m}, \qquad c_{sky} = \alpha 10^{-0.4 m_{sky}}
\end{equation}
where $\kappa, \alpha$ are quantities that can be written in terms of physical constants, physical parameters of the optical system, and noise equivalent area of the effective PSF ($\mathbf{FWHMeff}$ as listed in {\opsim} outputs) of the visit, all of which are known or measured quantites. Additionally, $\kappa$ depends on the total transmission function (optical system and atmosphere) through the throughput integral $T_b$ which changes from observation to observation, mostly driven by airmass and clouds), while $\alpha$ depends on the system transmission function through the system throughput integral $\Sigma_b,$ which is constant except for tiny differences caused by flexure of the system, or slowly over the years through the evolution of the system. The signal to noise ratio of the flux measurement is driven by the Poisson error due to both the source and sky counts. Since {\opsim} outputs do not contain $\kappa$ or $\alpha,$ but an equivalent set of variables, it is convenient to eliminate some of them in terms of quantities that are measured in a survey or available as simulated quantities in the {\opsim}  outputs like the five sigma depth $m_5,$ the sky brightness $m_{sky},$ and the PSF width provided in {\opsim} in terms of $\mathbf{FWHMeff}$. The general expression is 
\begin{equation}
    \kappa = \frac{25\times 10^{0.4 m_5}}{2}\left(1 + \sqrt{(1 + \frac{4 \alpha}{25} 10^{-0.4 m_{sky}})}\right)
    \label{eqn:general_counts}
\end{equation}
    which reduces to the familiar background dominated limit of $5 \sqrt{\alpha} \times 10^{0.2(2 m_5 - m_{sky})}$ in the limit where $\sqrt{c_{sky}} >> 1.$ This is similar in spirit in which $\sigma_{rand}$ is calculated in~\citet{overview}. These expressions relate $\kappa$
   to physical constants, physical parameters of the optical system that are constant in time through $\alpha,$ 
   and the quantities $m_{sky}, m_5, \mathbf{FWHMeff}$ available in {\opsim}. It should be remembered that all of these quantities $\alpha, m_{5}, m_{sky}$ are not independent, and therefore Eqn.~\ref{eqn:general_counts} does not imply that changing $\alpha$ by changing the PSF would change $\kappa$. 
On the other hand, if the small difference between $T_b$ and $\Sigma_b$ is ignored so that $\frac{\alpha}{\kappa}$ is considered to be a measured quantity from the measured PSF, one can find an expression for $\kappa$ in terms of the {\opsim} quantities $m_{sky}, m_5, \mathbf{FWHMeff}$ and the pixel size
\begin{equation}
    \kappa = \frac{25 \alpha}{\kappa} 10^{0.4(2m_5 - m_{sky})}\left( 1 + \frac{\kappa}{\alpha} 10^{-0.4(m_5 - m_{sky})} \right).
    \label{eqn:simlib_zpt}
\end{equation} 
without worrying about the physical characteristics of the optical system.

Thus, our goal is to obtain these terms for each visit in a transient light curve from the {\opsim} output. This is explained in a step by step procedure in SubSection.~\ref{ssec:objectives}

\subsection{Input Data: Operation Simulator Outputs}
To summarize the methodology used, we start by describing the input data product, namely the outputs from {\opsim}. 
The {\lsst} project simulates observing strategies using the Operations Simulator ({\opsim}) and the resulting sequence of pointings with properties of observations are disseminated in the form of a sqlite database. The database contains multiple tables, and the most important ones for our purpose are the `summaryAllProps' and `proposal'\footnote{In version 3 ,the `summaryAllProps' table was called the `summary' table.} The `proposal' table is a  table of scientific surveys or proposals, each of which have their own requirements in terms of desired visits and survey properties, along with a unique integer identifier `proposalId'. Currently, LSST has the Wide Fast Deep survey, a Deep Drilling Field survey, a Southern Galactic Cap Survey, a Milky Way Survey, and a Northern Ecliptic Spur survey in different geographical regions and different survey strategies applied to each of them. 

The `summaryAllProps' table is the sequence of simulated observations based on the simulated conditions throughout the ten year period. Each row of the table is an observation or a telescope pointing which we will refer to as `visits'. The row for a visit is identified by an integer `observationId` with important properties characterizing the observation as well as the `proposalId` whose criteria it satisfies. The characteristics of the observations include the pointing location, the time of observation, the bandpass in which the observation is made, the seeing and the PSF, the sky brightness, and the five sigma depth. The seeing is based on historical data, while the sky brightnesses are computed using a data-driven model \citep{2016SPIE.9910E..1AY}. Together, these two tables tell us about all of the simulated observations, and the scientific proposal or survey that they were taken to satisfy. These represent the sum-total of information available about the simulated strategies and are sufficient to generate catalog simulations. Complete details on such quantities are available from the schema of the output in the relevant version \footnote{\url{https://www.lsst.org/scientists/simulations/opsim/summary-table-column-descriptions-v335}}, \footnote{\url{https://lsst-sims.github.io/sims_ocs/tables/summaryallprops.html}}. In the current versions, the pointings are located on a discrete grid with an integer (fieldID) identifying each point on the discrete grid. There is no fundamental requirement that an observing strategy uses such a grid, and it is likely (and already true in some alternative simulators) that this grid does not exist; consequently the methodology we will describe below does not use this feature. To give an idea of the sizes involved, a typical operations simulator output contains about 2.5 million visits, while typical {\opsim} databases have a size of about 4.6 GB. 
There are some very specific details of {\opsim} outputs that are not obvious on first encounter. We attempt to list them here
\begin{itemize}
    \item Most of the proposals in the current baselines are non-overlapping. If there was a spatial location that was observed by survey WFD, it is not observed by a survey like Southern Celestial Pole or the Northern Ecliptic Spur. However, this is not true for WFD and DDF, and DDF fields can be observed by WFD as well. There is no reason that future mini-surveys will not have such overlapping properties.
    \item For a small fraction of cases, there can be multiple (actually two) rows of the summary table which point to the same visit. This happens whenever a particular visit satisfies the requirements of two different proposals or surveys. Currently, this is seen in the overlapping area of the Wide Fast Deep / Deep Drilling Field due to the previous point.
    \item While some outputs of the Operations Simulator come with a column of \verb ditheredRA   and \verb ditheredDec , these are added post-facto to the operation simulator output. Discussion of what the dithers should be is still ongoing, but it is useful to have the capability to replace these dithered observations with other dithers obtained from external sources. 
\end{itemize}  
\begin{figure}
    \begin{center}
        {\includegraphics[width=0.4\textwidth]{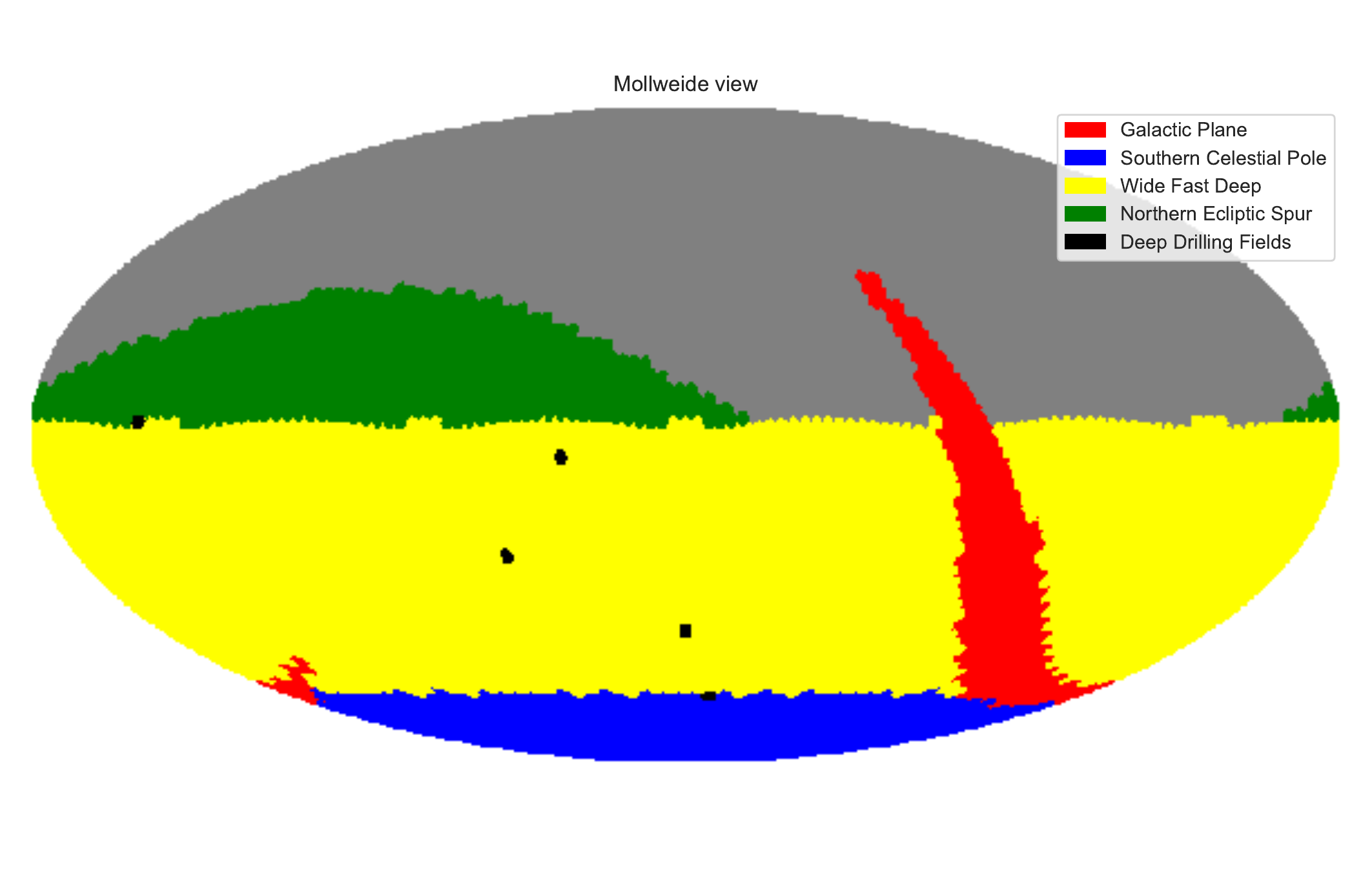}}
        \caption{An all-sky representation in Celestial coordinates in the Mollweide projection of the different {\lsst} proposals or surveys}
    \end{center}
    \label{fig:lsst_surveys}
\end{figure}
\subsection{Objectives}
\label{ssec:objectives}
\label{define_obs}
To further detail our objectives, we first define some terms that we will use in this paper. For any particular visit in LSST, a sky location within an angular radius of 1.75 degrees (the radius of the LSST focal plane) will be said to be \textbf{`observed by LSST during this visit'}. In reality, this is an approximation: LSST chips do not completely fill out the focal plane. There are parts of the circular disk that are not covered by the rectangular geometry of the chips, as well as chip gaps between the chips. Thus, the set of points observed by LSST during a visit according to the above definition is a superset of the points actually observed by the visit. We will ignore this distinction, except to note that the fill factor of chips is about \fillfactor\footnote{\url{https://www.lsst.org/about/camera/features}}.
Given a sequence of visits (or rows of {\lsst} {\opsim}  output) and a sky location, one can find the sequence of visits that will observe the sky location according to this definition. As this quantity will be used repeatedly in this paper, we will for brevity, refer to such a subset of all of the visits in an {\opsim} output as the `visit set' associated with a point on the sky. 


In terms of the terminology defined above, our objectives are quite simple: 
\begin{enumerate}
    \item Given an {\opsim} output, and a sky location in terms of Right Ascension (RA) and Declination (Dec),  we need a simple API to obtain the visit set of this location, i.e., the sequence of visits in the {\opsim} output that observe this location.
    \item \label{desiderata:large_comp} Since the {\opsim} outputs are large ($\sim 2.5\rm{~million}$ visits) and the number of {\transients} in LSST simulation volumes can be large ($\sim \rm{ ~millions}$) for abundant and bright {\transients} like SNIa, this could lead to ${\mathcal{O}}(10^{12})$ simple computations if done by brute-force in a naive way. We would like the process to be reasonably fast and not be a huge load on the memory requirements. Note, while the number of cosmologically useful SN in LSST will be smaller than the number of supernovae exploding in the observable volume, simulations have to simulate all of the supernovae before applying selection cuts to identify cosmologically useful supernovae.
    \item Pre-compute this information on a dense grid and serialize to {\snana} observation library formats to enable fast computations.
    \item \label{desiderata:version_stability} Since the Operations Simulation schema changes from version to version in terms of names, even though the conceptual setup remains the same, we would like to account for these changes and provide a stable interface for a catalog simulator.
\end{enumerate}

\section{Results}
\label{sec:results}
We present a simple, open source modular python package {\oss}  based on other open source libraries, particularly the package \verb Scikit-learn ~\citep{scikit-learn} to meet each of our objectives.
The code~\citep{rbiswas4_2019_2671955} is available online~\footnote{\url{https://www.github.com/lsstdesc/OpSimSummary}}, while the particular release described in this paper will be linked at the end.
While the actual implementations are somewhat different in terms of packages used, some of the key ideas are inspired by those used in \verb MAF  . We first explain how this code meets each of our objectives:

\subsection{Objective 1: API to collect visits observing a transient}
\label{ssec:api}
This package achieves our objective of collecting visits observing a transient. It takes the publicly available {\lsst}  project provided {\opsim}  outputs (in {\opsim} version 3 and 4, as well as the two other schedulers that were used: the Feature Based Scheduler~\citep{2018arXiv181004815N} and AltSched~\citep{2019arXiv190300531R}) as input, and provides an API for obtaining the visits for a point source at a sequence of arbitrary locations (defined by RA and Dec values). The code structure and examples for doing this are in the appendix of this paper, and available with the source code itself. It also allows for the usage of an additional set of dithers input as the filename of a file in Comma Separated Values (CSV) format. If the sources to be simulated can be simulated independently, distribution is trivial to achieve by splitting their locations into arrays and using these arrays independently.

\subsection{Objective 2: Computational Efficiency}
\label{ssec:comp_eff}
While the problem of enumerating all the transients, and the visits that observe each one of them is naively a ${\mathcal{O}}(N_{\rm{visits}}) \times {\mathcal{O}}(N_{\rm{transient}}),$ it is intuitively clear that an easier computation should be possible. Since one does not require the computation of distances to visit centers that are too far away the computation could take advantage of this.

There are different ways of implementing this intuitive idea of locality of visits. For example, a simple approach is choosing a convenient set of sky locations $pl$ at which the visit sets are actually computed and approximating the visit set of an arbitrary point (for example the set of point source locations $tl$) by the visit set of a deterministically selected grid point. Thus such schemes are defined by two components:
\begin{enumerate}
\item a selection of points $pl,$ at which the visit sets $v$ will be computed with no approximation. For the approximation to make a computing time difference, it would be nice for the size of $pl$ to be significantly smaller than the size of transients.
\item A mapping from the visit sets $v(x)$ for any point $x$ in $tl$ to the visit sets $v(y)$ of points $y$ in $pl$.
\begin{equation}
v(x) = v(\{v(y)\}), \qquad x \in tl, y \in pl.
\end{equation}
\end{enumerate}

A very simple algorithm along these lines would be nearest-neighbor-interpolation, where the component (2) would be defined by assigning to an arbitrary point $x \in tl$, the visit sets of the point in $pl$ closest to $x$.
Interpolation techniques exploit the smoothness of the function being interpolated. Here the `function' under consideration is a map which returns the visit set of a point. While observing conditions in the sky vary reasonably smoothly with location and time, the set of points being observed by a visit is determined by a hard boundary (edge of the focal plane). Any time such an edge falls between two points, one of the two points will be observed and the other will not. As the distance between two points decreases, the probability of such a visit also decreases, but for a large number of true visits in a visit set (in the WFD survey of {\lsst}, this is $\sim 1000$), this would still be expected to happen. This implies that despite the smoothness of observing conditions with spatial locations, the visit set associated with points would not be `interpolated' as well quantities like sky conditions. For a dense enough set of points, such a strategy could still provide an excellent approximation to the true visit sets. Of course, pre-computation of the quantities in a dense set and their storage could itself be challenging, particularly if several versions of survey strategies are analyzed.

An elegant way to exploit the locality of visits without using the smoothness of the visit set is the use of a Tree  data structure to partition the data based on spatial positions, so that we should expect a scaling of ${\mathcal{O}(N_{\rm{transient}})} \times {\mathcal{O}}(log(N_{\rm{visits}})).$  As far as the distance computations are concerned, ie. if we ignore the position of the chips etc., then {\it{this calculation does not involve any additional approximation, and the speed attained is simply due to an organization of the calculation}.}

Here, we use a  Tree  implementation to exploit the locality of visits and provide a simple API to compute the visit set associated with individual visits. This should be easy to use for a simulator in the sense described above. This is done by using an implementation within the package `Scikit-learn'~\citep{scikit-learn} called `BallTree'~\citep{sklearn_api}. We also use the API to pre-compute visit sets for a particular set of points to obtain approximate visit sets for each point, through an interpolation scheme for the well known {\snana} code as described in the next subsection. %
\subsection{Objective 3: {\snana} observation libraries}
For transient simulations, {\snana} has historically utilized the idea of splitting the sky to a relatively small set of pre-determined points. {\snana} simulates transients at only these locations. The abundance of transients simulated at each of these locations is tuned so that the expected number of transients (based on rates, survey volumes etc.) starting within any period of time over the total survey footprint is the sum of the number of transients during the same time period at these locations. To do such simulations, {\snana} reads in a pre-computed set of telescope pointings of a survey called `simlib fields' and the observing conditions associated with each pointing observing each of the simlib fields from an ASCII file known as a {\snana} observation library, with a specific format. An important objective of the {\oss} codebase is to provide precomputed observation libraries for {\snana} to enable simulations of {\lsst}. Previous versions of this codebase have been used to generate observation libraries used for {\snana} simulations and analyses in the {\lsst} DESC Science Requirement Document~\citep{2018arXiv180901669T}, while the code and features described here were primarily for the data generation of the {\plasticc} challenge~\citep{2018arXiv181000001T}, as described in the {\plasticc} model and simulations paper~\citep{2019arXiv190311756K}. We therefore include a script to use the more general API of ~\ref{ssec:api} to produce observation library files files which we are using for {\snana} simulations of {\lsst}. We proceed to describe the method by which such files were generated, by first describing the quantities being used by {\snana} and how they are related to {\opsim} quantities. We then describe the procedure we follow (in the script) to generate these observation library files: this includes the selection of footprints, selection of simlib fields, and then computing the quantities and writing them out.  

\begin{figure}
\begin{center}
\scalebox{1.}
{\includegraphics[width=0.5\textwidth]{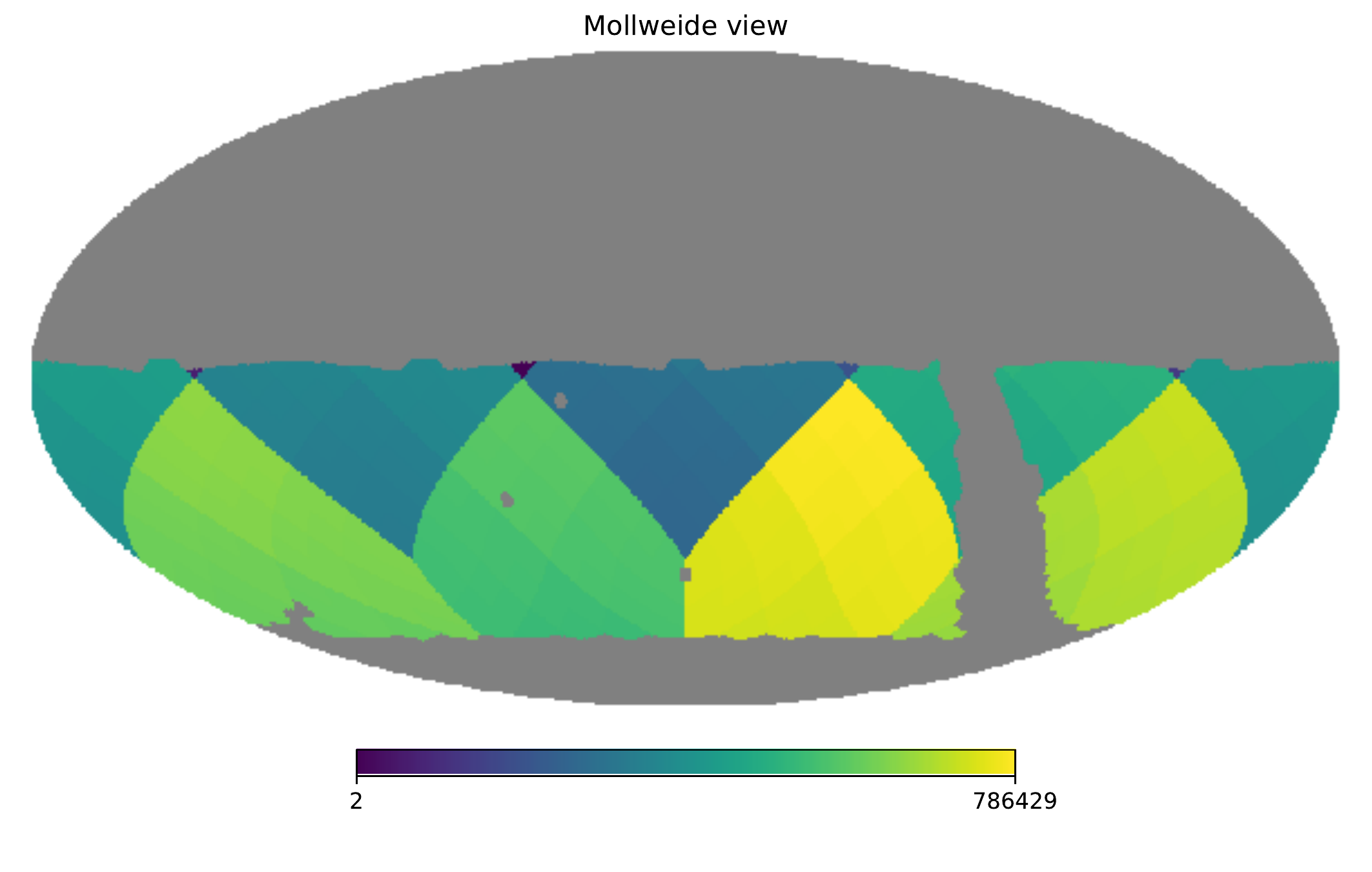}}
{\includegraphics[width=0.5\textwidth]{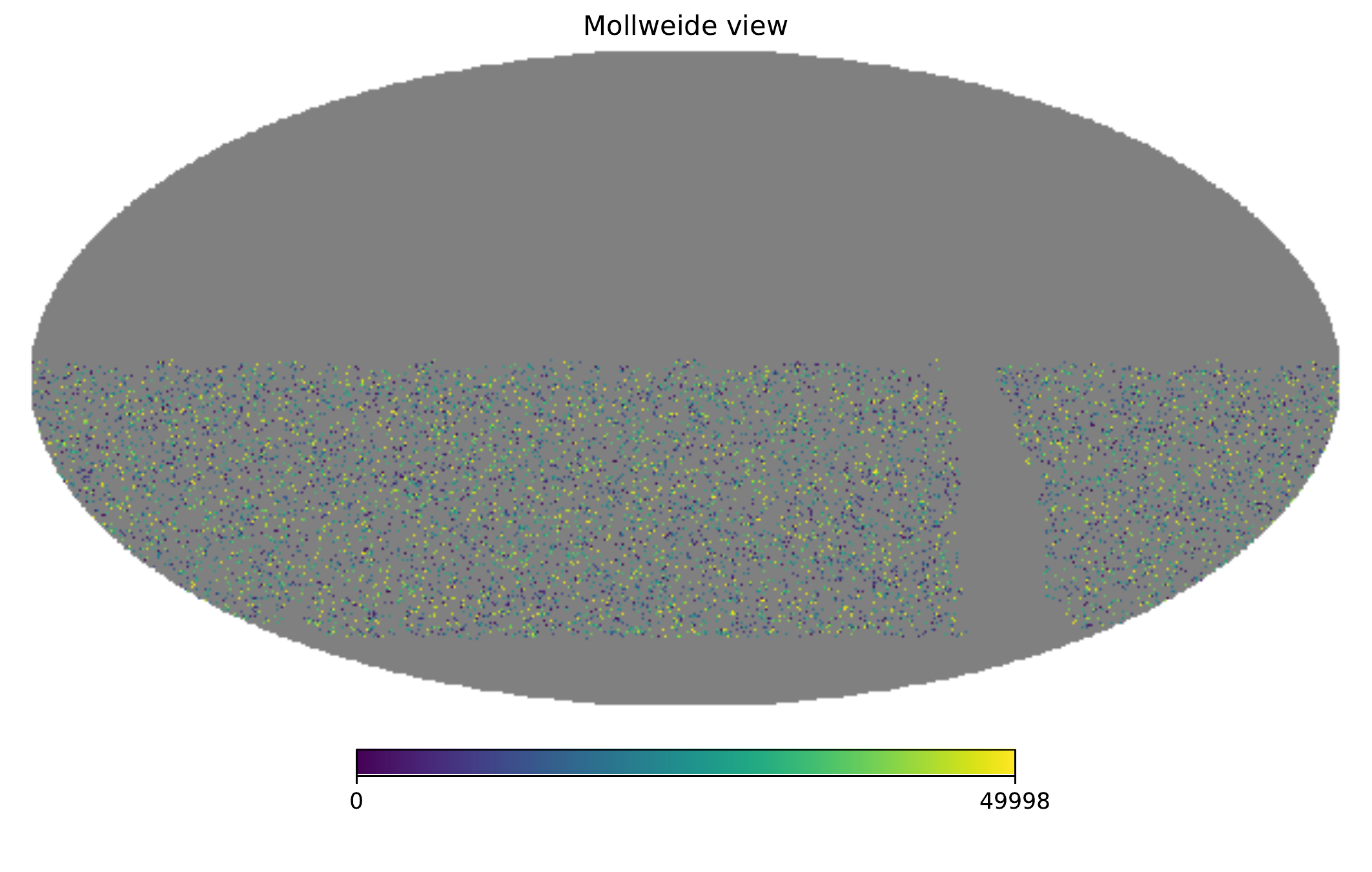}}
{\includegraphics[width=0.5\textwidth]{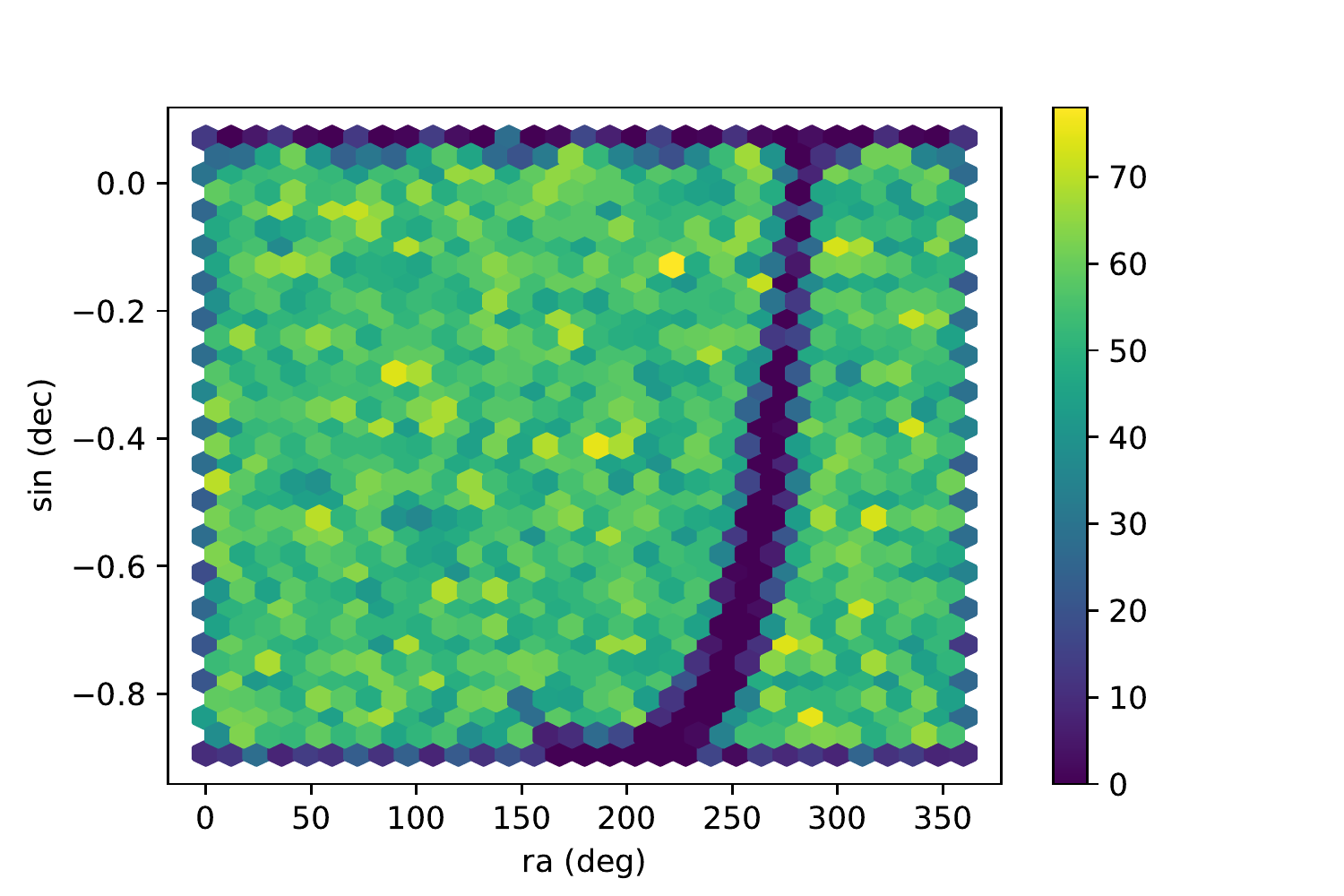}}
\caption{observation library fixed position choices: (Top Panel) Healpixels from NSIDE=256 (nest scheme) filled by the WFD survey used in observation libraries, which excludes the DDF areas as holes. The colors show the healpixel id, which for the particular NSIDE and scheme uniquely identify the healpixel. The solid color pattern shows that the healpixel ids have been written out in increasing order of healpixel ids. (Middle Panel) a sample of 50,000 healpixel positions, rather than all of the healpixel positions in a random order that is written to observation library files. here the color represents the order in which the healpixels are written to the observation library, and the lack of the solid color pattern shows that it is random rather than in increasing order of healpixel ids. (Lower Panel). A hex binned 2D histogram of the number of selected simlib fields in $RA$ and $\sin(Dec).$ Since hex bins in these transformed coordinates are of equal area, the color uniformity away from the footprint edges demonstrates uniform sampling to obtain the points.
\protect{\label{fig:simlib_checks}}}
\end{center}
\end{figure}

First we tie the quantities in the observation library file to {\opsim}  outputs, with a brief justification of the procedure. We then discuss the process of choosing the discrete locations at which these evaluations take place. The {\snana} observation library quantities (bold-faced on the left hand side of Eqn.~\ref{eqn:opsim2simlib}) are related to the {\opsim} quantities (bold-faced on the right hand side of Eqn.~\ref{eqn:opsim2simlib}) through simple transformations as:

\begin{equation}
\begin{gathered}
\label{eqn:opsim2simlib}
\begin{split}
\mathbf{PSF1} &= \mathbf{FWHMeff} / 2.35 / \mathrm{pixelSize} \\
A &= 1.51 \times \mathbf{FWHMeff} \\
\mathbf{ZPTAVG} &= ZPTApprox + ZPTCorr \\
ZPPTApprox &=  2.5\log_{10}(25A) + (2 \mathbf{m_5} - \mathbf{m_{sky}}) \nonumber\\ 
ZPTCorr   &= 2.5\log_{10}\left(1.0 + \frac{10^{-0.4(\mathbf{m_5} - \mathbf{m_{sky}})}}{A} \right) \nonumber\\
\mathbf{skySig} &= \mathrm{pixelSize} \times 10^{-0.4 (\mathbf{m_{sky}} - \mathbf{m_{5}})}\\
\mathbf{PSF2} &= 0 \\
\mathbf{PSF2/1} &= 0
\end{split}
\end{gathered}
\end{equation}

In particular, the variables $\mathbf{PSF1, PSF2}$ of {\snana}  meant to describe the PSF shape are represented by a simple two dimensional isotropic Gaussian profile with a radial standard deviation. Since $\mathbf{FWHMeff}$ is related to the effective PSF area in {\opsim}, we set $\mathbf{PSF2, PSF2/1}$ to zero, and then the quantity $\mathbf{PSF1}$ is simply related to $\mathbf{FWHMeff}$ of {\opsim} through the first equation of Eqn.~\ref{eqn:opsim2simlib}.  The quantity $\mathbf{skySig}$ of {\snana}  is related to $\frac{\alpha}{\kappa}$ of
Eqn.~\ref{appendixeqn:alphaoverkappa}. Finally, {\snana} uses the approximation of $\alpha/ \kappa =1$ in Eqn.~\ref{eqn:simlib_zpt}. Thus, one can see that $\mathbf{ZPTAVG} = 2.5\log_{10}{(\kappa)}$ from Eqn.~\ref{eqn:simlib_zpt}, with $\alpha/\kappa$ set to unity.

Next, we discuss the selection of discrete points which we will refer to as `simlib fields', where the visit sets, and the quantities above are calculated for each visit in the visit sets using the API of ~\ref{ssec:api}. This is done by first selecting the DDF and WFD footprints, followed by uniformly selecting points from each of these footprints. The first step in this procedure is the selection of the DDF footprint. The footprint selection is done by going over all the visits in the {\lsst} DDF minisurvey, which are identified by a `proposalId` index in {\opsim}. We tesselate the sky into small {\healpix} {\footnote{\url{https://healpix.sourceforge.io}}} pixels~\citep{2005ApJ...622..759G} which we shall refer to as healpixels. We tesselate the sky with healpixels with ($\mathrm{NSIDE}=256$, equivalently of pixel area $\approx$~$0.05$~square degrees, which is roughly the size of a {\lsst} chip), and find the healpixels that contain at least one point which is observed by at least a threshold number of visits. These healpixels together make up the footprint of the DDF minisurvey. The provided script arguments allow the user to set the threshold, but the default is $500$ visits over ten years. As the healpixels have equal area, the total area of the DDF footprint is the number of healpixels multiplied by the area of each healpixel.

The footprint of the WFD survey is also found in a similar way. First healpixels belonging to the DDF are removed, and any other healpixel that contains at least a single point which has a visit set with a number of visits over a threshold are taken to be the WFD footprint. This gives us the footprint of the WFD without any point which has been observed by DDF (ie. with holes around the DDF location) as shown in the top panel of Fig.~\ref{fig:simlib_checks}. The color shows the healpix id in the HEALPix nest scheme. Again the threshold can be defined by the user, but the default value used is $500$. It should be noted that, estimates of the WFD area in the {\lsst} ~literature follow a different convention: the DDF area is not removed from the WFD footprint as we have done, and areas for the WFD footprint are often quoted with a threshold of $825$ visits, the median requirement ({\lsst} SRD) of WFD visits in a field.

We treat these two geographical areas as different surveys whose areas have been measured. In order to simulate observation library files for each of these surveys, we first choose a fixed number of simlib fields within the survey footprint. This fixed number can be chosen by the user, and the default number is 50,000 points for the WFD. The default number for the DDF is 150. The numbers are roughly proportional to the survey area footprints found above. The simlib fields are chosen randomly from the footprint. Here, random implies that any area of a fixed size within the footprint has the same probability of having a certain number of points. However, uniformly sampling an odd-shaped footprint is somewhat complicated. While the codebase can uniformly sample healpixels (using rejection sampling), and therefore the footprint which is made of healpixels, this is inherently slow. So, we choose to not worry about it. Instead we pick a random integer from the set of healpixel IDs for the healpixels in the footprint, giving us different healpixels with equal probability. This is sufficient for our requirements of uniform sampling, but would not pass specific tests of isotropy, as the points are chosen to be healpixel centres. The uniformity of sampling is shown in the lower panel of Fig.~\ref{fig:simlib_checks} where a hex-binned plot of the selected points against $RA$ and $\sin(Dec),$ where the color scale shows the number of selected points in each bin is shown to be roughly uniform, excluding the location of the Milky Way which is not observed in detail by the WFD and DDF surveys in the current strategy.

For each of the selected simlib fields, we use the API of subsection.~\ref{ssec:api} to obtain the visit sets observing these points from both WFD and DDF proposals of the {\opsim} output. By construction, the visit set of points in the DDF footprint includes visits from both WFD and DDF proposals of {\opsim}, which is the correct way to simulate transients. Points on the WFD footprint have visit sets that contain visits only from the {\opsim} WFD proposal. We calculate the derived quantities required for {\snana}  using the quantities available from {\opsim} through Eqns.~\ref{eqn:opsim2simlib} as described above, and write out the information in the format  required by {\snana}  to simlib files. We have found that when used with {\snana} for rare transients, it is important to randomize the ordering in which these healpixel position is read (ie. not according to increasing healpixel id) and randomize the order of writing the selected points out. An example of such a selected sample, with the colors showing the serial ordering of these points is shown in the middle panel of Fig.~\ref{fig:simlib_checks}. Often, to speed up the simulation process, and control sizes of outputs, these simlibs are co-added over nights by {\snana}, before simulation.

Finally, in the bottom panel, we check that the distribution of points is truly uniform (barring anomalous regions like the Milky Way where there are no WFD/DDF visits, by checking the rough uniformity of the hexbin plot in the bottom panel of Fig.~\ref{fig:simlib_checks}.

We end this subsection on {\snana} simlibs with a description of a few generic features of the simlibs for the current baseline cadences of {\lsst}. The sky area of the WFD and the DDF footprints calculated in the method described above are $\sim 18.0 \times 10^3$ and $\sim 47.6$ sq. deg. respectively. These footprints are modelled by $50, 000$ and $150$ simlib fields respectively, resulting in an average area per simlib field of $0.36 $ and $0.32$~sq. deg. respectively. Without compression, these ASCII files have sizes of about $4.6$ and $0.3$ GB respectively.

\subsection{Validation, Performance and Accuracy}
\begin{figure}
\begin{center}
\hspace*{-0.5cm}
\scalebox{1.}{\includegraphics[width=0.5\textwidth]{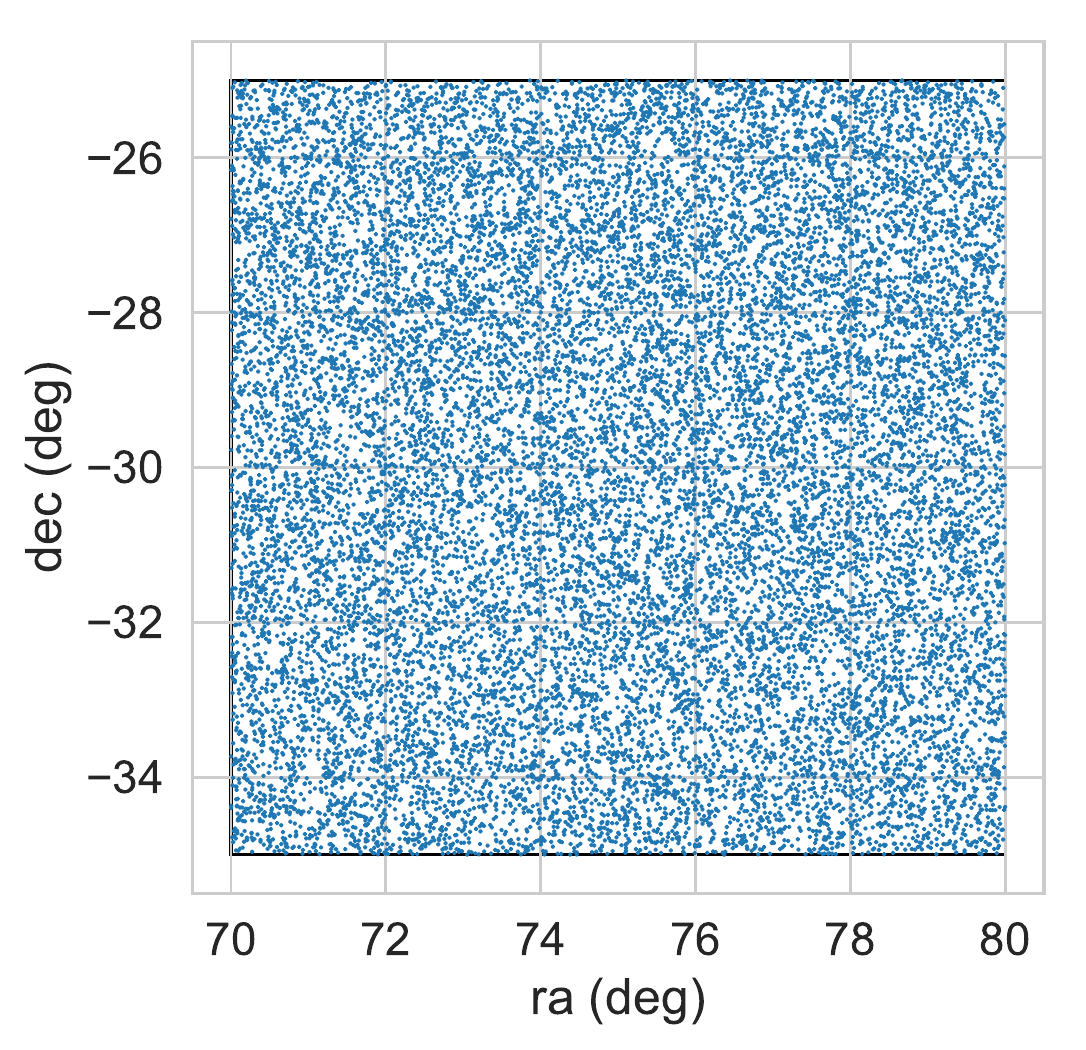}}
\caption{The location of the sky patch and a scatter plot of the points chosen uniformly in the area to evaluate the accuracy of discretization schemes \protect{\label{fig:samples}}
}
\end{center}
\end{figure}

First, as part of a standard test, we check that the visit sets from the API match the values with the naive solution. To give an idea of the time required in the current setup,  (after a common initializaton for all sources which mostly involves reading in the database) the code required $37.2 \pm 3.1$ sec to obtain visit sets for $50,000 $ sky locations spread over the same $100$~sq. deg~patch of the sky as shown in Fig.~\ref{fig:samples}. 

As we have seen, approximating the visit list by the visit list of a nearby pre-computed point is a useful approach, not only because it enables the use of other software, but because during the actual simulation obtaining the visit list is almost instantaneous.  We have noted that this will inevitably result in differences with the correct calculations, but the the approximation approaches the correct results as the set of points where the pre-computation is performed is made denser. Making the set of pre-computed points arbitrarily dense requires pre-computations and storage. So, quantifying a relationship between the denseness of points (eg. simlib fields) and the accuracy of a pre-computations at those points helps understand the tradeoffs. With the tools available here, we can quantify this accuracy with density.

\begin{figure}
\begin{center}
\hspace*{-0.5cm}
\scalebox{1.}{\includegraphics[width=0.5\textwidth]{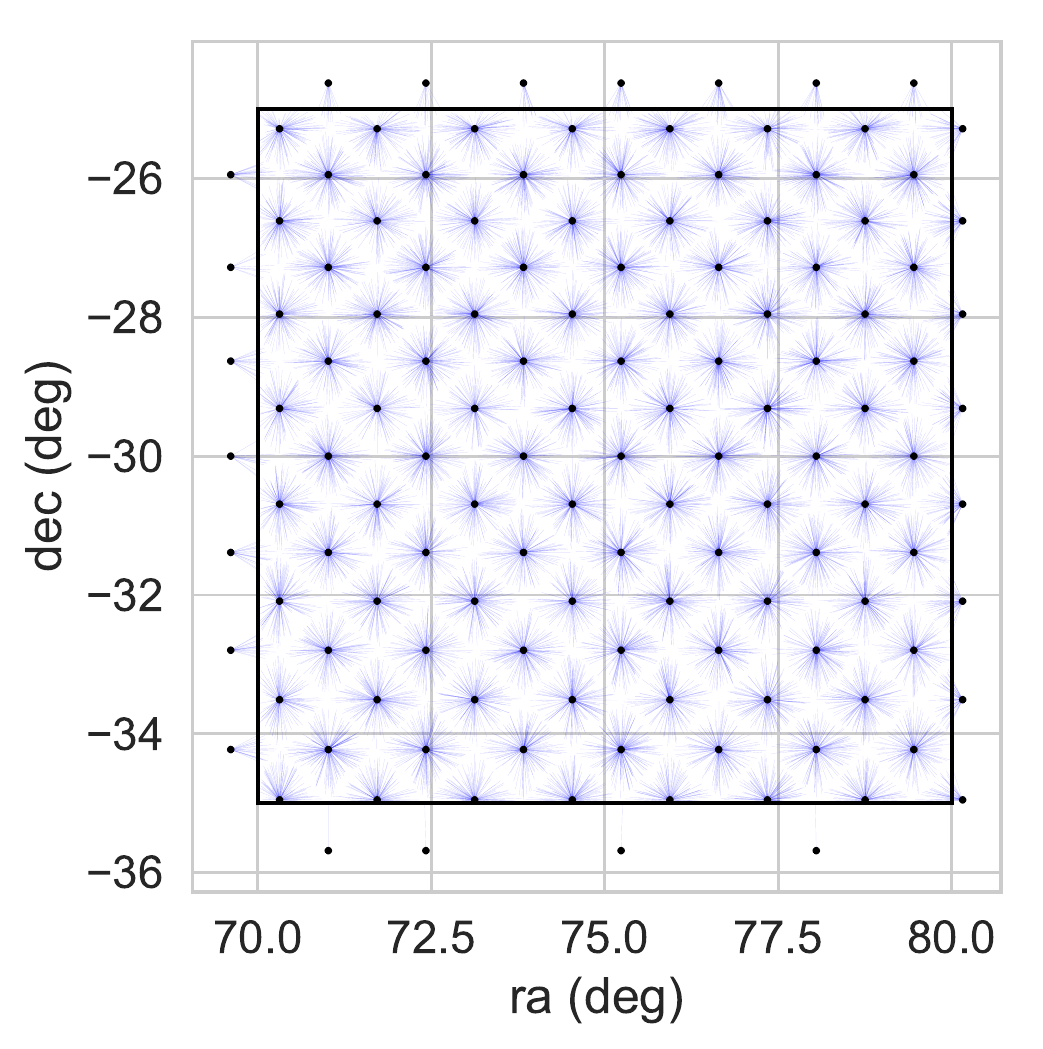}}
\caption{Scatter plot of the points (black dots) at which the visit sets are computed for a scheme where $NSIDE = 64.$ Since the scheme results in assigning to any point $x$ in Fig.~\ref{fig:samples} the visit sets of exactly one black point $y$ of this scatter plot, we use the blue lines to connect the pair of points ($\{x, y\}$ to show the mapping operation. \protect{\label{fig:displacement}}
}
\end{center}
\end{figure}

\begin{figure}
\begin{center}
\scalebox{1.}{\includegraphics[width=0.5\textwidth]{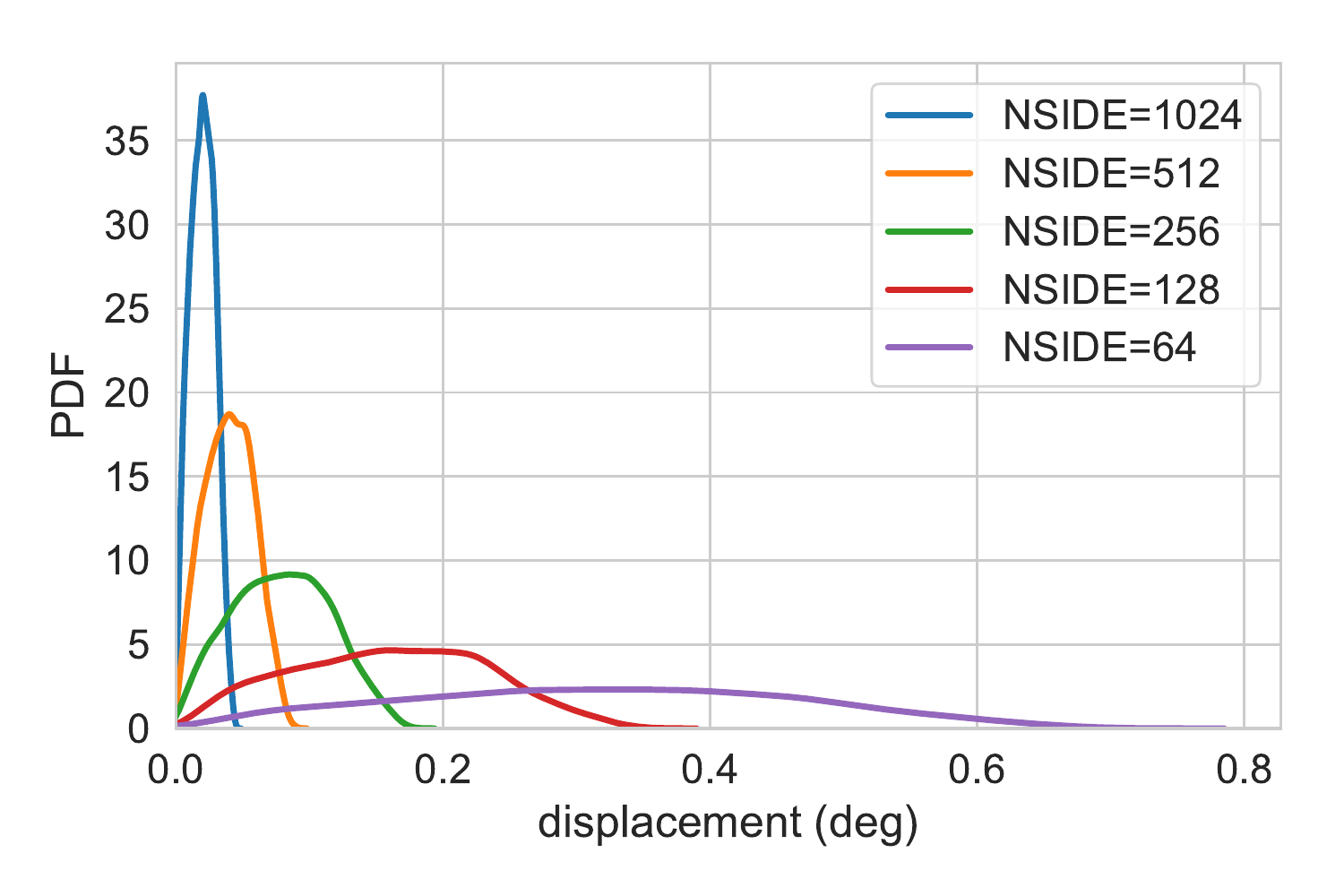}}
\caption{Distribution of lengths of the blue lines in Fig.~\ref{fig:displacement} in degrees connecting a point $x$ in Fig.~\ref{fig:samples} and a black point in Fig.~\ref{fig:displacement} where its visit set is evaluated. Shorter displacements indicate better approximations.
\protect{\label{fig:length_dist}}
}
\end{center}
\end{figure}

\begin{figure}
\begin{center}
\scalebox{1.}{\includegraphics[width=0.5\textwidth]{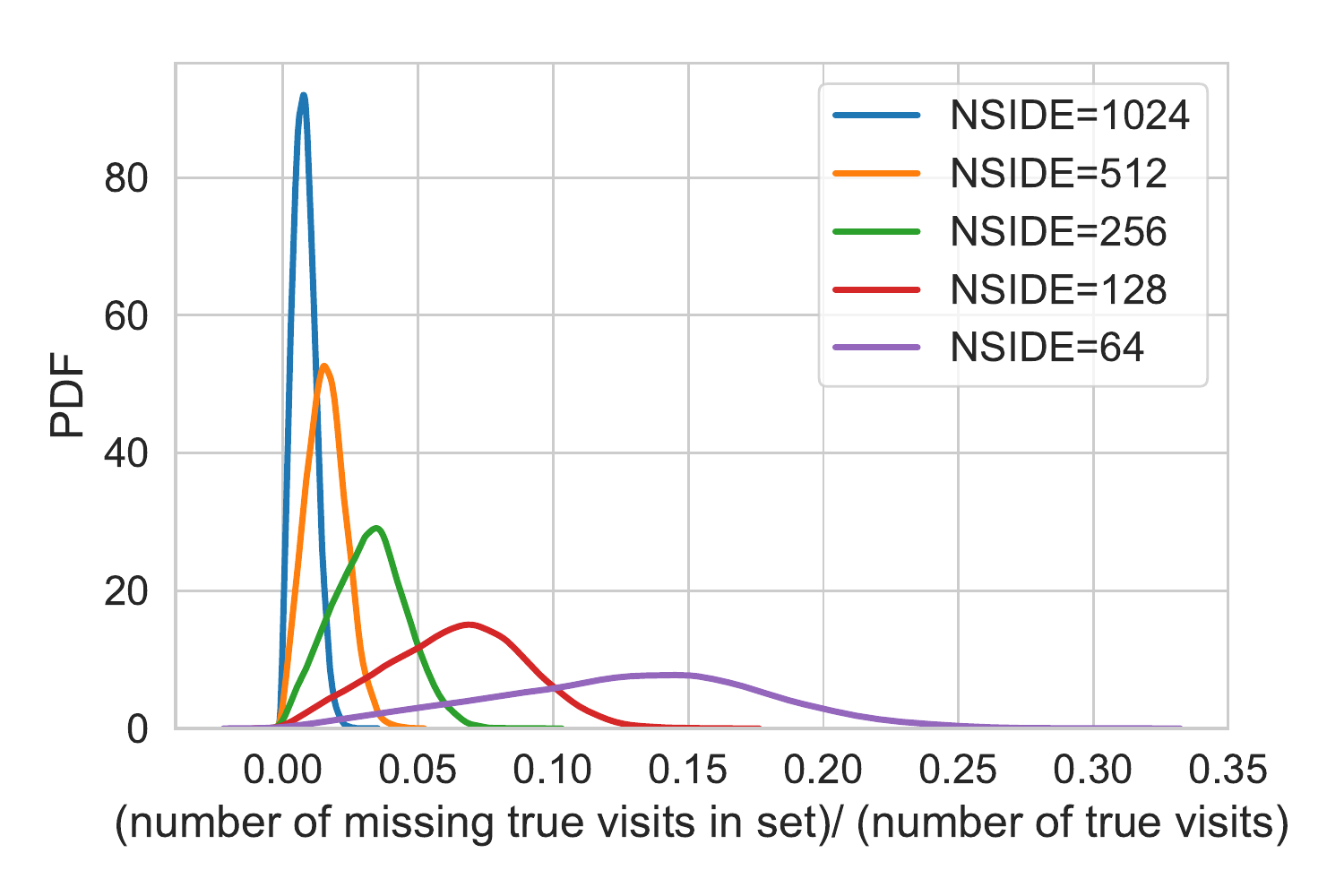}}
\caption{Distribution of the ratio of the number of visits missing in the approximate computation of visit sets to the true number of visits computed at points in Fig.~\ref{fig:samples}
\protect{\label{fig:true_visits}}
}
\end{center}
\end{figure}

\begin{figure}
\begin{center}
\scalebox{1.}{\includegraphics[width=0.5\textwidth]{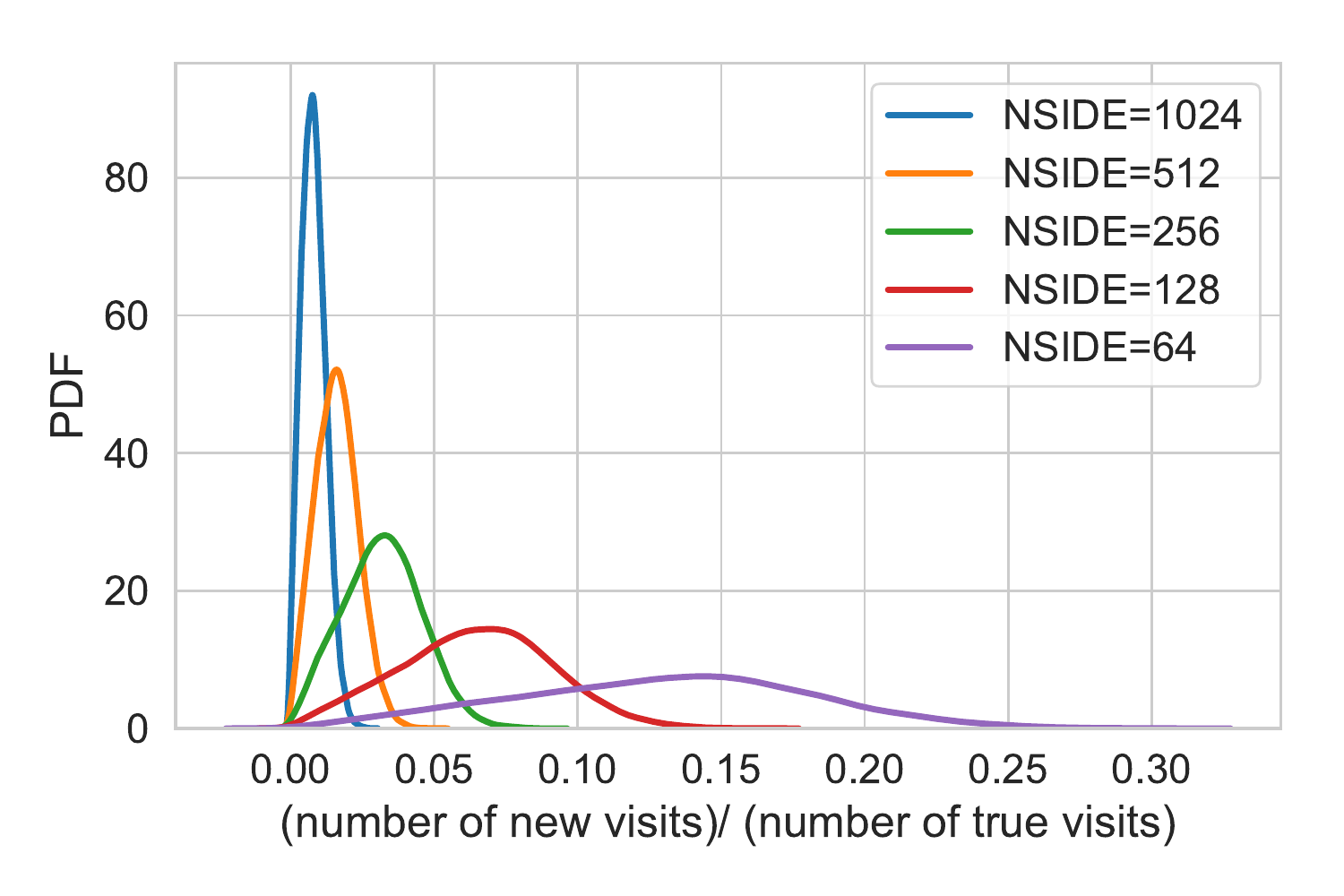}}
\caption{Distribution of the ratio of the number of visits in the approximate computation that are not in the true visit sets to the number of visits in the true visit sets
\protect{\label{fig:newvisits}}
}
\end{center}
\end{figure}

\begin{figure}
\begin{center}
\scalebox{1.}{\includegraphics[width=0.5\textwidth]{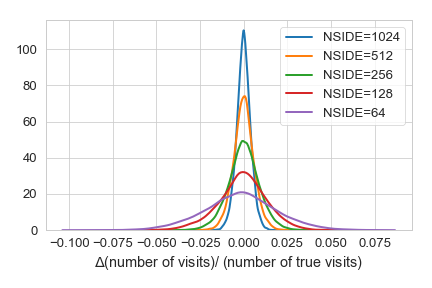}}
\caption{Distribution of the ratio of the difference in the number of visits in the approximate computation and the number of visits in the true visit set to the number of visits in the true visit sets
\protect{\label{fig:diffpdf}}
}
\end{center}
\end{figure}
To quantify the accuracy of visit sets in different approximation schemes with pre-computed points, we choose a discretization scheme: following the discussion of such schemes, this can be described by the two components (a) a pre-determined $pl$ at which visit sets $tv(p)$  for $p \in pl$ are actually computed, and (b) a prescription to assign visit sets using 
$v(x) = v(\{p, tv(p)\}$ for any point $x$. 

Since we expect the accuracy of a discretization scheme to depend strongly on the density of
points $pl$ at which the visit sets were actually computed, and perhaps weakly, on the actual discretization scheme, we choose a particular, convenient and relevant (as the use of such schemes are already prevelant) kind of discretization scheme and vary the density. 
To that end, we set up the following exercise: we choose a patch of the sky in the $RA$ range of $(70,80)$ degrees, and the $Dec$ range of $(-35,-25)$ degrees, and select a sample $tl$ of  $\rm{size} = \nsize $ points uniformly within this area. This patch of the sky and a scatter plot of these points is shown in Fig.~\ref{fig:samples}.  The points chosen for precomputation $pl$ are the positions of Healpixels near the sky patch at different values of \verb NSIDE  = $\{64,~128,~256,~512,~1024\}.$ We remind the reader that Healpixels of a particular \verb NSIDE  all have equal area, and $12 \times \rm{NSIDE}^{2}$ Healpixels tile the surface of the entire sphere. This allows the calculation of the healpixel areas in square degrees, or their `resolution' which is simply the square root of the pixel area. Accordingly, these pixels have areas of $0.84, 0.21,  0.05, 0.01, 0.003$ square degrees and resolutions of $55.0{\arcmin},~27.4{\arcmin},~13.7{\arcmin},~6.9{\arcmin},~3.4{\arcmin}.$ This is explained in Fig.~\ref{fig:displacement}, where we show the setup for the choice of $NSIDE = 64.$ Here, the black points show the $pl, $ the healpixel centers at which we actually compute the visit sets for this scheme, while the blue lines connect pairs of points $\{x, y\}$ where $x \in tl$ is a point in Fig.~\ref{fig:samples}, and $y \in pl$ is a black point in Fig.~\ref{fig:displacement} where the visit sets assigned to $x$ are the visit sets computed at $y$. Before looking at the results, we recall from the discussion that we expect the accuracy of visit sets to be related to the length of these blue lines in Fig~\ref{fig:displacement} connecting the points $x, y,$ and therefore show the distribution of such lengths in Fig.~\ref{fig:length_dist}.

We compute the true visit sets for each of the points in $tl$ shown in Fig.~\ref{fig:samples}. We then use our discretization scheme for different values of \verb NSIDE , to compute only the visit sets at the precomputed points $pl,$ and assign to each point in $tl$ the visit sets of the healpix position of the Healpixel where the point lies. We refer to this as the approximated visit set under the specific discretization scheme. Following previous discussion, we know that the inaccuracies will result in (a) visits in the original visit set being missing in the approximated visit set, and (b) new visit sets in the approximate visit set that do not exist in the true visit set. While we would like to keep both of these quantities small, we also recognize that they will grow with the size of the true visit set. Hence, the appropriate quantity to monitor is the ratio of the number of missing visits to the number of true visits, and the ratio of the number of new visits to the number of true visits. In Fig.~\ref{fig:true_visits}, we show the distribution of the number of missing visits to the number of true visits for different values of $NSIDE$. The plot shows that the distribution is quite broad for $NSIDE = 64,$ which has a resolution of $\sim 54{\arcmin},$ and peaks at about $15 \%$ of the visits missing, while for $NSIDE = 1024,$ which has a resolution of $\sim 3 {\arcmin},$ this distribution is very narrow, and peaks at slightly lower than a percent of missing visits, with the values in between following the trend in both the width of the distribution and the location of the peak. The distribution of the ratio of new visits to the true visit sets is also shown in Fig.~\ref{fig:newvisits}. The distribution of these visits is quantitatively very similar to the visits in Fig.~\ref{fig:true_visits}.  

We should note that the difference between the total number of visits in the approximated visit set  and the true visit set is statistically smaller, as these two errors affect the size in the opposite direction. However, replacing the visits by a different set of visits does not preserve the time of observation and can have significant differences for example, due to differences of bright and dark time. It also does not necessarily preserve the bandpass of observation. We know that the distance estimates of supernovae Type Ia are closely linked to band coverage, and thus cases where the missing visits correspond to the less frequently observed bands in {\lsst}, it is likely that the additional visits are going to be in the more frequently observed bands. Such differences might be important for science programs like supernova cosmology, even though investigating these details is beyond the scope of this work. With this cautionary note, in Fig.~\ref{fig:diffpdf} we show the ratio of differences in sizes between the approximate visit set and the true visit set. Given the quantitative similarity between Fig.~\ref{fig:newvisits} and Fig.~\ref{fig:true_visits}, it is not surprising to see that the distribution of the difference in the number of visits normalized by the number of visits in the true visit set is centred at zero with a width decreasing with $NSIDE$ from a few percent to $\lesssim 1 \%.$

\section{Summary and Discussions}
In this paper, we discuss the importance of catalog simulations of Time Domain Sources (TDS) for the study of analysis methods, and survey strategy of {\lsst}. Survey strategies of {\lsst} are currently simulated by the LSST project using {\opsim}; such simulated survey strategies are made public in the form of sqlite databases that are outputs of {\opsim}. We discuss the transformations of the set of quantities in {\opsim} that are required for catalog simulations. We also discuss the usefulness of re-ordering the outputs in terms of {\opsim} visits observing a particular sky location, delineating the necessity of such an API. While conceptually simple, we discuss why a naive solution is inefficient, particularly during the simulation of abundant sources. As strategies to address this issue, we discuss exploiting the locality of visits using a Tree data structure; and approximating the problem by serializing pre-computed results for use with a simulator. This strategy makes the step during simulations essentially instantaneous, but inevitably results in errors which can be minimized by choosing a very dense set of pre-determined points at the cost of large file sizes.

We present an open source modular python source software package for such operations, which contains an API for reading in {\opsim} outputs and re-ordering them to obtain the visits for each point. Thus, a simulation code can directly use this API to obtain the important quantities. A Tree is used to speed up the calculations. We also use the obtained visits, along with simple transformations of {\opsim} quantities to serialize the results for a set of points in the form of an {\snana} simlib. The script to perform this is also made available as part of the {\oss} package. Currently {\oss} works with {\opsim} outputs of version 3, and 4, along with outputs of Feature Based Scheduler and AltSched. 

We study the accuracy of the approximate pre-computed visit sets as a function of the density (or average separation) of the points at which the visit sets are actually computed, and show that at large average separations between these points, the visit set of sky locations have several visits missing, while several new visits not originally in the visit set are inserted. According to the numbers calculated for the current strategies, we would expect the the current method to include $\sim 10 \%$  visits are missing while a similair number of $\sim 10 \%$ visits that were not in the true visit set were added.

This code has been used through the direct use of API in the study of serendipitous discoveries of Kilonovae using the {\lsst}~\citep{2018arXiv181210492S} which also formed part of a {\lsst} DESC survey strategy white paper for Wide Fast Deep Fields in {\lsst}~\citep{2018arXiv181200515L}. {\snana} observation library  files~\citep{biswas_rahul_2017_1006719} generated through previous versions of {\oss} (and distributed publicly with the {\snana}  code) have been used in the study of serendipitous detection of Kilonovae~\citep{2018ApJ...852L...3S} and the LSST DESC Science Requirement Document~\citep{2018arXiv180901669T}. This paper describes the improved versions of {\snana} observation library files (simlibs) currently available, developed primarily for the data generation of {\plasticc}  ~\citep{2018arXiv181000001T}, as described in the {\plasticc} models and simulations paper~\citep{2019arXiv190311756K}. These observation library files have also been used in the supernova simulations using {\snana}  used for the supernova cosmology analyses in the LSST DESC Survey Strategy white papers~\citep{2018arXiv181200515L,2018arXiv181200516S}.

\subsection*{Acknowledgments}
RB would like to thank David Cinabro for sharing his DACG project at the beginning of this work, Rick Kessler for stimulating discussions, and particularly on its use with respect to {\snana} and Lynne Jones for help in understanding MAF.

This paper has undergone internal review in the LSST Dark Energy Science Collaboration. 
The internal reviewers were Philippe Gris and Isobel Hook, and the authors would like to thank them for their comments.
Author contributions are listed as follows.
\textbf{RB}: Initiated and led project, wrote the {oss} package, drafted paper, derived results.
\textbf{SFD}: Provided support for middleware connecting OpSim with simulations of astrophysical sources.
\textbf{RH}: Code beta testing, discussed results.
\textbf{AGK}: Motivated the project and consulted with RB particularly in its initial phases.
\textbf{PY}: Supported OpSim and MAF usage.\\
During this work, RB was partially supported by the Washington Research
Foundation Fund for Innovation in Data-Intensive
Discovery and the Moore/Sloan Data Science Environments
Project at the University of Washington and the Swedish Research Council (VR) through the Oskar Klein Centre. RB was further supported by the research environment grant ``Gravitational Radiation and Electromagnetic Astrophysical Transients (GREAT)" funded by the Swedish Research council (VR) under Dnr 2016-06012. \software{Aside from the standard python package, this work used the following software packages: numpy~\citep{2011CSE....13b..22V}, healpy~\citep{Zonca2019} and {\healpix} packages~\citep{2005ApJ...622..759G}, pandas~\citep{mckinney-proc-scipy-2010}, sqlalchemy, scikit-learn~\citep{scikit-learn,sklearn_api}, while the examples use Jupyter Notebooks~\citep{Kluyver2016JupyterN}}


The DESC acknowledges ongoing support from the Institut National de Physique Nucl\'eaire et de Physique des Particules in France; the Science \& Technology Facilities Council in the United Kingdom; and the Department of Energy, the National Science Foundation, and the LSST Corporation in the United States.  DESC uses resources of the IN2P3 Computing Center (CC-IN2P3--Lyon/Villeurbanne - France) funded by the Centre National de la Recherche Scientifique; the National Energy Research Scientific Computing Center, a DOE Office of Science User Facility supported by the Office of Science of the U.S.\ Department of Energy under Contract No.\ DE-AC02-05CH11231; STFC DiRAC HPC Facilities, funded by UK BIS National E-infrastructure capital grants; and the UK particle physics grid, supported by the GridPP Collaboration.  This work was performed in part under DOE Contract DE-AC02-76SF00515.




\bibliography{main,lsstdesc}
\appendix
\section{Point Sources and SNR}
\label{appendix:snr}
Given the physical parameters describing a telescope, and a description of the sky and astrophysical sources, one can calculate quantities like the expected number of photons collected from a point source in the sky (ie. no background galaxy), or the sky. Combining this with observing conditions based on seeing, airmass etc., one can calculate a good estimate of the expected signal to noise ratio of an observation. We follow the discussion in~\cite{LSE-40}, keeping the gain $g=1$ in our calculation (For an extensive discussion including latest updates to LSST values, see ~\cite{jones_2016_192828}).

For a point source with intensity $\epsilon(\lambda)$ as a function of its wavelength $\lambda,$ the number of photons collected with an exposure time $T$ in a telescope with collecting area $A$ is given by:
\begin{equation}
c_{source} = \frac{AT}{h} \int_{0}^{\infty} F_{\nu}(\lambda) \lambda^{-1} d{\lambda} S^{tot}(\lambda)
\end{equation}
where the flux density $F_{\nu}(\lambda)$ is the frequency derivative of the intensity 
$F_{\nu}(\lambda) \equiv \frac{d\epsilon(\lambda)}{d\nu},$ 
while $S^{tot}(\lambda)$ is the total transmission probability due to the atmosphere, the telescope system, and $h$ is the Planck constant. We note that $S^{tot}(\lambda)$ is also a function of time through the dependence of the atmospheric transmission functions on airmass, and atmospheric conditions.

Similarly, using the intensity per unit area of the sky $b_{\nu}(\lambda),$ one can calculate the time averaged number of photons collected in $n_{eff}$ pixels as
\begin{equation}
    \frac{c_{sky}}{n_{eff}} = \frac{AT}{h} \int_{0}^{\infty} b_{\nu}(\lambda) \lambda^{-1} d{\lambda} S^{sys}(\lambda) pa
\end{equation}
where $pa$ is the area of a pixel. 
In order to estimate the number of photons collected from the source and sky during a particular exposure from he observed pixel counts, one uses estimators such as `aperture photometry' and `psf photometry'. 
In each of these, one can use a value of $n_{eff}$ pixels based on the observing conditions. 
For the estimator used in PSF photometry, this is given by 
\begin{equation}
    \label{appendixeqn:radialgaussianpsf}
    n_{eff} = 2.27 \frac{FWHM}{pixelScale}^2
\end{equation} if the PSF profile is assumed to be a single radial Gaussian.

These counts obviously depend on the flux densities in exactly the same way as magnitudes in the bands, and so be calculated just by knowing the source magnitudes and the sky brightness in $mags/{arcsec}^2,$ without requiring complete information on the flux densities. 
\begin{eqnarray}
    m_{source} &=& -2.5 \log_{10}\left(\frac{\int_{0}^{\infty} F_{\nu}(\lambda) \lambda^{-1} d{\lambda} S^{Total}(\lambda)}{\int_{0}^{\infty} F^{std}_{\nu}(\lambda) \lambda^{-1} d{\lambda} S^{Total}(\lambda)}\right) \\
    c_{source} &=& 10^{-0.4m_{source}} \int_{0}^{\infty} F^{std}_{\nu}(\lambda) \lambda^{-1} d{\lambda} S^{Total}(\lambda) \frac{AT}{h} \\
               &=& 5328 \left(D/6.43m\right)^2 \left(T/30s\right)T_b \times 10^{-0.4 m_{source}}
\end{eqnarray}
where the numerical values in the last line assumes that the magnitude is in the AB system (ie. $F^{std}_{\nu} = 3631 Jy$), and that the area is a circular disk of diameter $D,$ and the throughput integral
$T_b$ is 
\begin{equation}
    T_b \equiv \int_{0}^{\infty} S^{tot}(\lambda) \lambda^{-1} d\lambda.
\end{equation}
One can do a similar calculation for the counts of sky photons: 
\begin{eqnarray}
    m_{sky} &=& -2.5 \log_{10}\left(\frac{\int_{0}^{\infty} b_{\nu}(\lambda) \lambda^{-1} d{\lambda} S^{sys}(\lambda)}{\int_{0}^{\infty} F^{std}_{\nu}(\lambda) \lambda^{-1} d{\lambda} S^{sys}(\lambda)}\right) \\
    c_{sky} &=& 10^{-0.4m_{sky}} \int_{0}^{\infty} F^{std}_{\nu}(\lambda) \lambda^{-1} d{\lambda} S^{Sys}(\lambda) \frac{AT\times pa \times n_{eff}}{h}\\ 
            &=& 5328 \left(D/6.43m\right)^2 \left(T/30s\right) \left(\frac{pa \times n_{eff}}{0.04}\right)\Sigma_b \times 10^{-0.4 m_{sky}}
\end{eqnarray} 
where the system throughput integral $\Sigma_b$ is 
\begin{equation}
    \Sigma_b \equiv \int_{0}^{\infty} S^{sys}(\lambda) \lambda^{-1} d\lambda.
\end{equation}

Hence, we see that we can write the photons counts as 
\begin{equation}
c_{source} = \kappa 10^{-0.4 m_{source}} \qquad c_{sky} = \alpha 10^{-0.4 m_{sky}}
\label{appendixeqn:prop}
\end{equation}
where we can write $\kappa, \alpha$ in terms of physical quantities emphasizing the fact that $T_b(t)$ changes with time, as does
$n_{eff},$ but is related directly to quantities that are supplied by most surveys (and in {\opsim}).
\begin{eqnarray}
    \kappa  &=& 5328 \left(D/6.43m\right)^2 \left(T/30s\right)T_b(t) \\ 
    \alpha  &=& 5328 \left(D/6.43m\right)^2 \left(T/30s\right) \left(\frac{pa \times n_{eff}(t)}{0.04}\right)\Sigma_b 
\end{eqnarray}

So that we get 
\begin{equation}
    \label{appendixeqn:alphaoverkappa}
        \frac{\alpha}{\kappa} =\frac{pa\times n_{eff}}{0.04}\left(\frac{\Sigma_b}{T_b}\right)
\end{equation}

The Signal to Noise Ratio (SNR) of a measured source can be found from
Poisson statistics:
\begin{equation}
    \mathrm{SNR} = \frac{c_{source}}{(c_{source} + c_{sky})^{1/2}}
    \label{appendixeqn:SNR_eqn}
\end{equation}
In practice, there may be other small sources of uncertainty such as read noise or other systematic errors
that could in principle be grouped together with the Poisson Noise in the
denominator of Eqn.~\ref{appendixeqn:SNR_eqn}.
Plugging Eqn.~\ref{appendixeqn:prop} into Eqn.~\ref{appendixeqn:SNR_eqn}, we can get
\begin{equation}
    \mathrm{SNR} = \frac{\kappa 10^{-0.4 m_{SNR}}}{(\kappa 10^{-0.4 m_{SNR}} + \alpha 10^{-0.4 m_{sky}})^{1/2}}
\end{equation}
If values of $m_{sky}$ and $m_5$ are supplied for an ovservation for a survey (as they often are), one can 
solve this to obtain Eqn.~\ref{eqn:general_counts}

\end{document}